\documentclass[prb,preprint,showpacs,preprintnumbers,amsmath,amssymb]{revtex4}
\usepackage{times}
\usepackage{mathptmx}
\usepackage{graphicx}
\usepackage{dcolumn}
\usepackage{bm}

\bibpunct{}{}{,}{s}{}{}

\begin{document}
\preprint{APL}

\title{Ion implanted Si:P double-dot with gate tuneable interdot coupling}

\author{V.C. Chan}
 \email{victor.chan@student.unsw.edu.au}
\author{T.M. Buehler}
 \altaffiliation[Now at ]{ABB Switzerland, Corporate Research.}
\author{A.J. Ferguson}
\author{D.R. McCamey}
\author{D.J. Reilly}
 \altaffiliation[Now at ]{Dept. Physics, Harvard University}
\author{A.S. Dzurak}
\author{R.G. Clark}

\affiliation{
Centre for Quantum Computer Technology, Schools of
Physics and Electrical Engineering \& Telecommunications,
The University of New South Wales, NSW 2052, Sydney Australia\\
}

\author{C. Yang}
\author{D.N. Jamieson}

\affiliation{ Centre for Quantum Computer Technology, School of
Physics, University of Melbourne, VIC 3010, Australia\\
}

\date{\today}


\begin{abstract}
We report on millikelvin charge sensing measurements of a silicon
double-dot system fabricated by phosphorus ion implantation. An
aluminum single-electron transistor (SET) is capacitively coupled
to each of the implanted dots enabling the charging behavior of
the double-dot system to be studied independently of current
transport. Using an electrostatic gate, the interdot coupling can
be tuned from weak to strong coupling. In the weak interdot
coupling regime, the system exhibits well-defined double-dot
charging behavior. By contrast, in the strong interdot coupling
regime, the system behaves as a single-dot.

\end{abstract}

\maketitle


Double-dot systems hold significant interest for both classical
\cite{lent1993} and quantum information processing
\cite{loss1998}. For quantum information processing (QIP),
semiconductor quantum double-dots with discrete states have proven
to be an excellent platform for research into quantum states and
their interactions \cite{hayashi2003,gorman2005}. In the use of
classical double-dots which have a continuous density of states,
applications have been demonstrated in single-electron logic
\cite{ichan2003} and quantum cellular automata
\cite{orlov1997,amlani1999} (QCA). A key aspect of both classical
and quantum double-dot systems is the interdot coupling, which
determines the strength of the interaction between the two dots
\cite{wiel2003}. We present a double-dot system, implemented in
silicon, with electrically tuneable interdot coupling.
Compatibility with silicon CMOS techniques is an advantage for
application in classical electronic circuits. In addition, silicon
is an attractive material for realizing a spin-based quantum
computer architecture for QIP \cite{kane1998} due to the long spin
coherence times associated with phosphorus doped silicon
\cite{tyryshkin2003}.

\begin{figure}
\includegraphics[width=6.5cm]{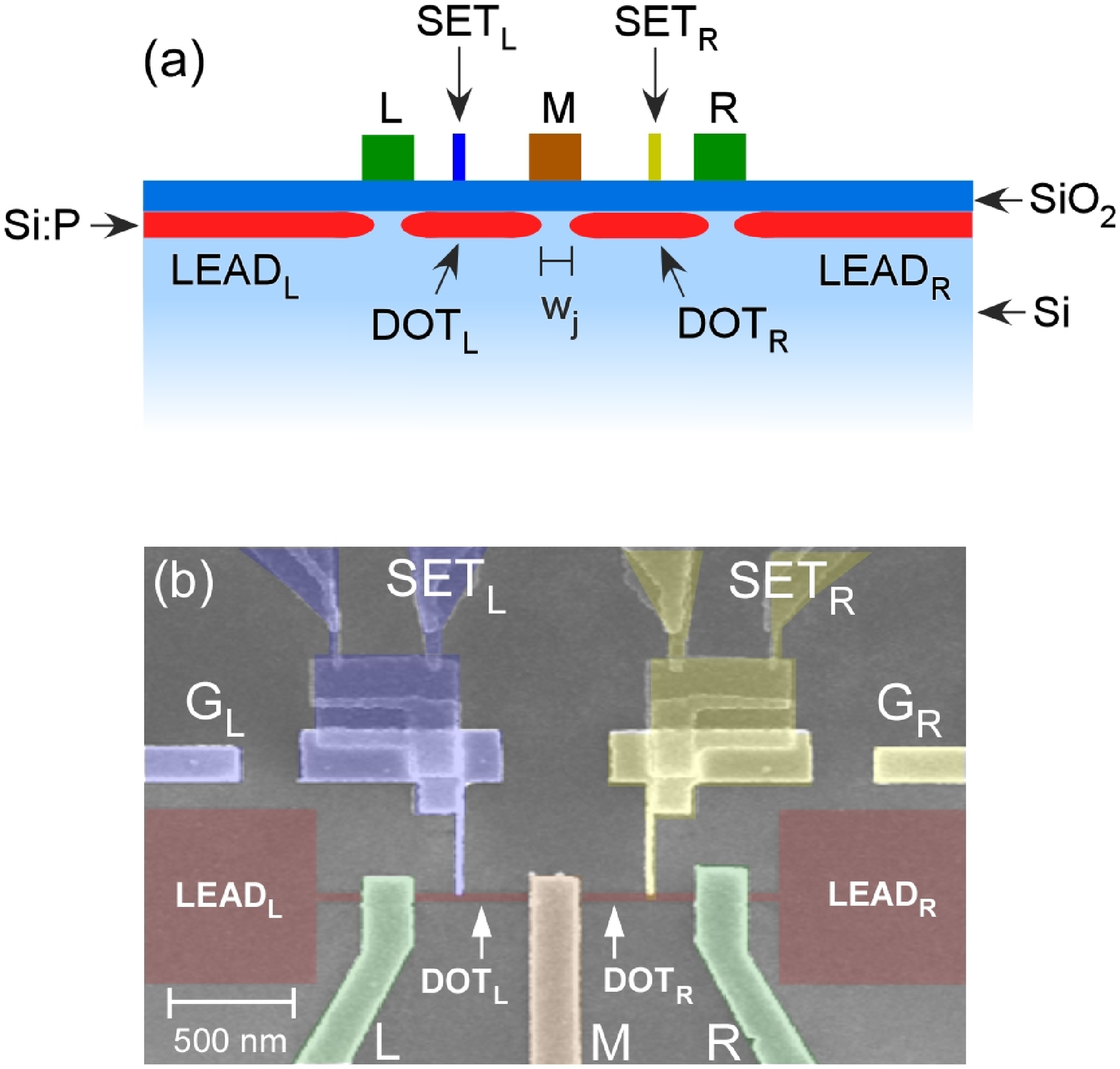}
\caption{\label{fig:one}}
\end{figure}

The silicon double-dot device presented here comprises two Si:P
dots and leads, with a metallic density of states, that are formed
by low energy (14 keV) phosphorus ion implantation. Ion
implantation provides a controllable method for creating
well-defined locally doped regions in silicon capable of
exhibiting single-electron charging \cite{chan2005siset}. The
tunnel junctions are formed by the undoped silicon regions and
have the form of nanoscale MOSFETs \cite{sanquer2000}, and a
surface gate provides control of the interdot coupling (Fig.
\ref{fig:one}(a)). Aluminum single-electron transistor (SET)
charge sensors coupled to the Si:P dots provide a means of
non-invasively detecting the charge state of the system. Charge
sensing is particularly relevant for device regimes not accessible
with direct transport measurements, for example for weakly coupled
and few-electron quantum dots
\cite{dicarlo2004,elzerman_nature04}.



The device investigated was fabricated on high resistivity ($>$ 5
k$\Omega$.cm) n-type silicon with a thermally grown 5nm gate
oxide. A 150 nm thick polymethylmethacrylate (PMMA) resist
patterned by e-beam lithography was used as a mask for ion
implantation. The dimensions of the dots were designed to be 70 nm
$\times$ 500 nm and the device was fabricated with tunnel junction
widths w$_{\mathrm{j}}$ $\simeq$100 nm (Fig. \ref{fig:one}(a)).
Phosphorus ions at 14 keV were implanted with an areal dose of
approximately $1.22$ $\times$ $10^{14}$ cm$^{-2}$ and a mean
implantation depth of $\sim$20 nm. A rapid thermal anneal (RTA)
was performed to repair implantation damage and electrically
activate the phosphorus dopants. Surface control gates were
fabricated by e-beam lithography in PMMA resist and TiAu
metallisation. The aluminum SETs were fabricated by double angle
evaporation with an \textit{in situ} oxidation using a shadow mask
formed in bilayer e-beam resist.

Figure \ref{fig:one}(b) shows an SEM image of a device similar to
that measured. SET$_{\mathrm{L}}$ and SET$_{\mathrm{R}}$ serve as
charge sensors for the charge states of DOT$_{\mathrm{L}}$ and
DOT$_{\mathrm{R}}$, respectively. Surface gates $L$ and $R$
control the electrostatic potential of DOT$_{\mathrm{L}}$ and
DOT$_{\mathrm{R}}$. Gate $M$ is positioned over the tunnel
junction between the two dots and is used to control the interdot
coupling. Gates $G_{L}$ and $G_{R}$ are used to position the SETs
to operating points of high charge sensitivity.


Electrical measurements were carried out at $T$ = 50 mK in a
dilution refrigerator. No magnetic field was applied to the device
($B$ = 0 T) and the SETs were operated in the superconducting
regime with a dc source-drain voltage applied to bias the SETs to
points of high charge sensitivity. All measurements were performed
in the linear transport regime of the double-dot system
\cite{wiel2003}, i.e. double-dot V$_{\mathrm{SD}}$ = 0 mV. SET
conductances were measured simultaneously using separate lock-in
amplifiers, with an excitation voltage of 20 $\mu$V at frequencies
$<$ 200 Hz.

\begin{figure}
\includegraphics[width=8.5cm]{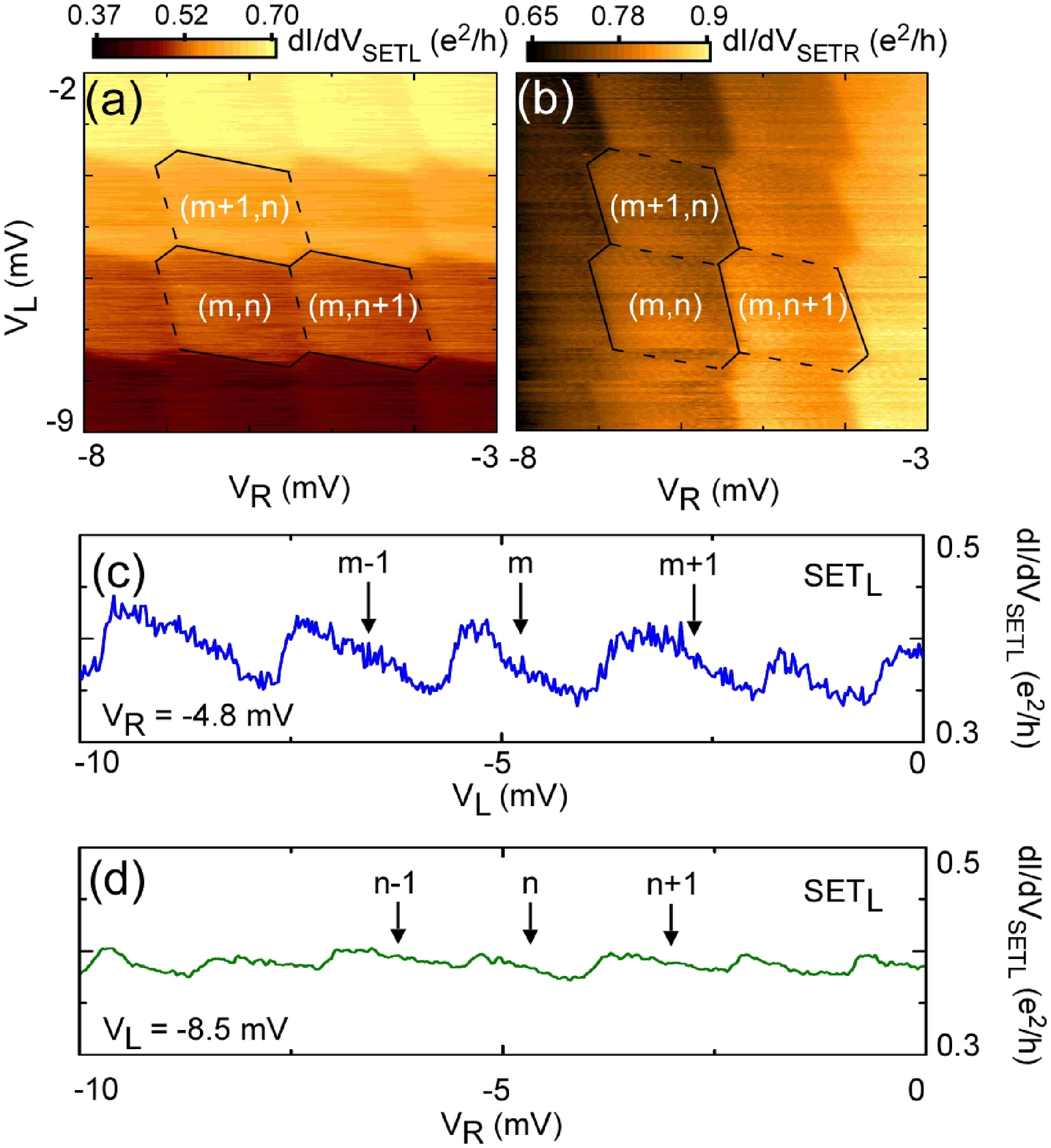}
\caption{\label{fig:two}}
\end{figure}

Figures \ref{fig:two}(a) and (b) show SET$_{\mathrm{L}}$ and
SET$_{\mathrm{R}}$ differential conductances measured as a
function of V$_{\mathrm{L}}$ and V$_{\mathrm{R}}$, with
V$_{\mathrm{M}}$ = 0 V. The charge stability diagram of a
double-dot system is given by hexagon shaped cells as a function
of gate voltage V$_{\mathrm{L}}$ and V$_{\mathrm{R}}$, each of
which represent a particular charge configuration \cite{wiel2003}.
The charge occupancies of DOT$_{\mathrm{L}}$ and
DOT$_{\mathrm{R}}$ are given by $m$ and $n$, respectively. The
horizontal (vertical) lines in the charge stability diagrams
represent LEAD$_{\mathrm{L(R)}}$ $\leftrightarrow$
DOT$_{\mathrm{L(R)}}$ transitions, where the charge occupancy of
DOT$_{\mathrm{L(R)}}$ is changed by one electron for adjacent
hexagons, e.g. ($m$, $n$) $\rightarrow$ ($m+1$, $n$). Triple
points occur when three different charge configurations are
degenerate, and interdot charge transitions (DOT$_{\mathrm{L}}$
$\leftrightarrow$ DOT$_{\mathrm{R}}$) take place across the
intersections of nearest neighbor triple points, e.g. ($m+1$, $n$)
$\rightarrow$ ($m$, $n+1$). The direct capacitances of gates $L$
and $R$ to their respective dots were determined from the period
of the LEAD$_{\mathrm{L(R)}}$ $\leftrightarrow$
DOT$_{\mathrm{L(R)}}$ charge state transitions to be
C$_{\mathrm{L,DOTL}}$ $\simeq$ 100 aF and C$_{\mathrm{R,DOTR}}$
$\simeq$ 80 aF, while the cross-capacitances were found to be
negligible. The difference in the direct capacitance values arise
from a slight misalignment between the gate and dots.

Single traces were taken from the data in Fig. \ref{fig:two}(a),
and SET$_{\mathrm{L}}$ differential conductance plotted as a
function of gate $L$ (Fig. \ref{fig:two}(c)) and gate $R$ (Fig.
\ref{fig:two}(d)). The `sawtooth' behavior in Figs.
\ref{fig:two}(c) and (d) is a signature of charge state
transitions as measured by charge sensing. There is a steady
polarization as a function of gate voltage and then a
discontinuity which signifies the detection of an electron
tunnelling event. Note that the sawtooth signal arising from
charge transitions in DOT$_{\mathrm{L}}$ has a larger amplitude
than the signal arising from charge transitions in
DOT$_{\mathrm{R}}$. Although each SET is primarily coupled to one
of the dots, a cross-capacitance between the SETs and their
opposite dots results in small induced charge on the SET islands.
The charge induced on the SET islands for each of the different
charge state transitions was evaluated and the values are shown in
Table \ref{table:one}. The magnitude of the charge induced on the
SET islands differs for each charge transition due to the
different capacitances associated with each of the tunnel barriers
in the device.

\begin{figure}
\includegraphics[width=8.5cm]{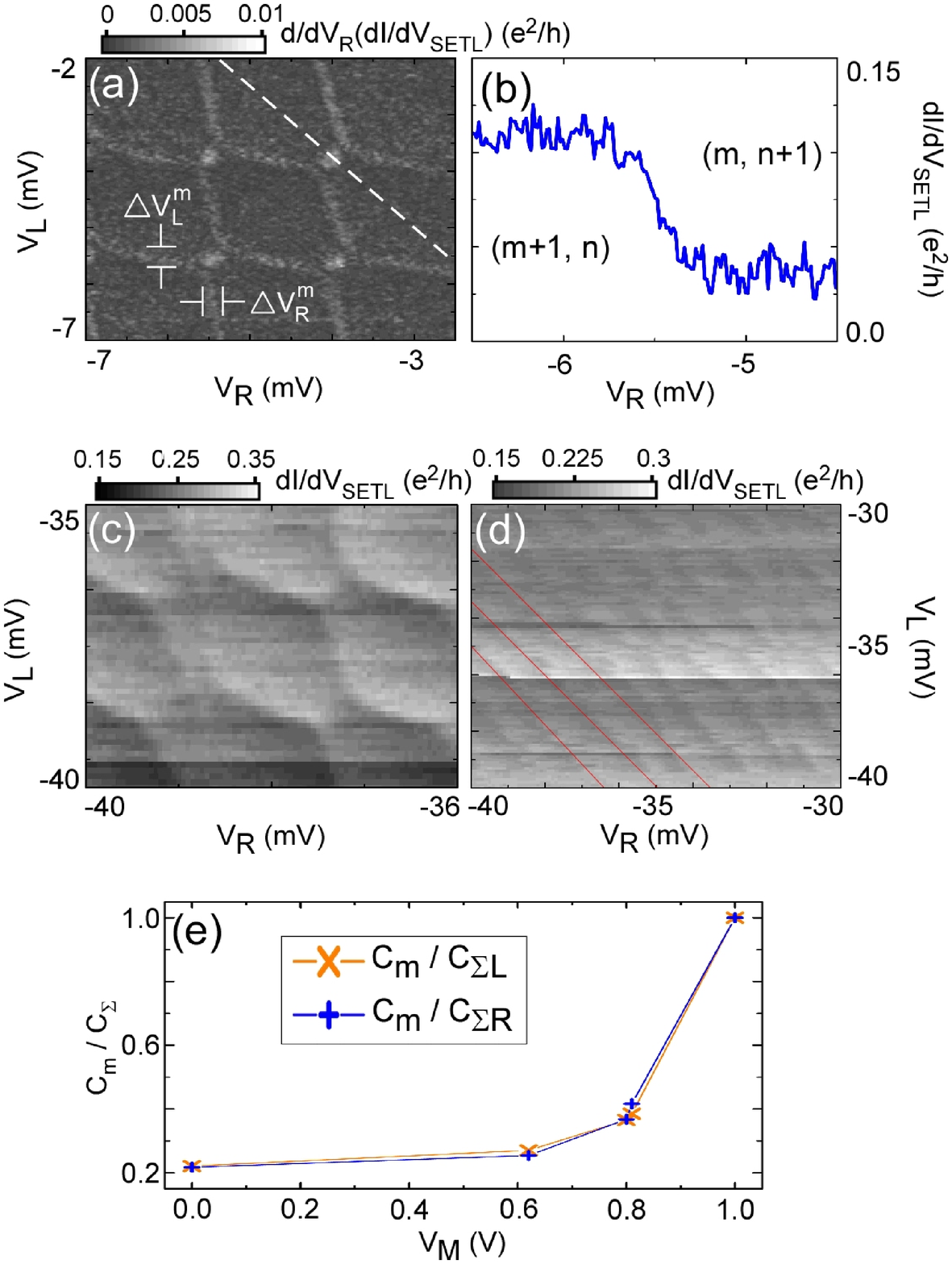}
\caption{\label{fig:three}}
\end{figure}

For interdot charge state transitions, the total charge of the
double-dot system remains fixed, i.e. ($m$ + 1, $n$)
$\leftrightarrow$ ($m$, $n$ + 1). Figure \ref{fig:three}(a) shows
a plot of SET$_{\mathrm{L}}$ differential conductance which has
been differentiated with respect to V$_{\mathrm{R}}$, enabling the
triple points can be resolved more clearly compared with Fig.
\ref{fig:two}(a). Taking a slice through the charging diagram as
indicated by the diagonal line, Fig. \ref{fig:three}(b) shows the
differential conductance of SET$_{\mathrm{L}}$ corresponding to an
interdot transition ($m$ + 1, $n$) $\rightarrow$ ($m$, $n$ + 1).
The interdot transition has the form of a Coulomb `step', and the
width of the step is dependent on the interdot coupling. The
interdot coupling between the two dots is characterized by the
mutual capacitance \cite{wiel2003} (C$_{\mathrm{m}}$) and the
tunnel coupling, given by the tunnel conductance \cite{waugh1996}
(G$_{\mathrm{int}}$). Conductance measurements of single tunnel
junctions (MOSFETs) indicate that at V$_{\mathrm{M}}$ = 0 V the
double-dot is in the weak tunnel coupling regime
(G$_{\mathrm{int}}$ $\ll$ 2e$^{2}$/h), so C$_{\mathrm{m}}$ was
treated as the dominant parameter determining the interdot
coupling. The electrostatic coupling C$_{\mathrm{m}}$ was
extracted from the triple point separation for a range of values
of V$_{\mathrm{M}}$.


The ratio of mutual capacitance C$_{\mathrm{m}}$ to total dot
capacitance C$_{\Sigma L(R)}$ was determined from the distance
between neighboring triple points by the equation
C$_{\mathrm{m}}$/C$_{\Sigma L(R)}$ =
$\triangle$V$^{m}_{\mathrm{R(L)}}$(C$_{\mathrm{R(L)}}$/e), where
$\triangle V^{m}_{\mathrm{R(L)}}$ is the triple point separation
in V$_{\mathrm{R}}$(V$_{\mathrm{L}}$) space \cite{wiel2003} (Fig.
\ref{fig:three}(a)). From Fig. \ref{fig:three} (a), the ratios
were evaluated to be C$_{\mathrm{m}}$/C$_{\Sigma L}$ $\sim$ 0.221
and C$_{\mathrm{m}}$/C$_{\Sigma R}$ $\sim$ 0.217 when
V$_{\mathrm{M}}$ = 0 V. The individual dot charging energies were
not determined for the device by a finite bias measurement,
however, the interdot coupling ratios indicate nearly identical
charging energies for both dots which is expected from the device
geometry. Figure \ref{fig:three}(c) show SET$_{\mathrm{L}}$
differential conductance, as a function of V$_{\mathrm{L}}$ and
V$_{\mathrm{R}}$, where V$_{\mathrm{M}}$ = + 0.81 V. Increasing
the positive voltage applied to gate $M$ increases the interdot
coupling, which is resolved as an increase in the separation
between the triple points. For V$_{\mathrm{M}}$ = + 0.81 V, the
interdot coupling ratios were evaluated to be
C$_{\mathrm{m}}$/C$_{\Sigma L}$ $\sim$ 0.385 and
C$_{\mathrm{m}}$/C$_{\Sigma R}$ $\sim$ 0.416.

As V$_{\mathrm{M}}$ is further increased, the interdot coupling
increases such that C$_{\mathrm{m}}$/C$_{\Sigma L(R)}$
$\rightarrow$ 1 (Fig. \ref{fig:three}(d), V$_{\mathrm{M}}$ = + 1.0
V). The device no longer exhibits the hexagon cells of a
double-dot charge stability diagram, but rather a series of
parallel lines associated with single-dot charging. Each line
corresponds to a transition in the charge occupancy of the
`single-dot' where the total charge is given by $m$ + $n$. Figure
\ref{fig:three}(e) shows the interdot coupling ratios
C$_{\mathrm{m}}$/C$_{\Sigma L(R)}$ as a function of
V$_{\mathrm{M}}$. Conductance measurements of single tunnel
junctions (MOSFETs) indicate that the tunnel junction between
DOT$_{\mathrm{L}}$ and DOT$_{\mathrm{R}}$ remains in the weakly
tunnel coupled regime, i.e. G$_{\mathrm{int}}$ $\ll$ 2e$^{2}$/h,
when V$_{\mathrm{M}}$ = + 1.0 V. This suggests that the increasing
interdot coupling is dominated by an increase in the mutual
capacitance C$_{\mathrm{m}}$ rather than an increase in the tunnel
coupling G$_{\mathrm{int}}$ \cite{livermore1996}. An explanation
for this is the formation of an electron accumulation layer
between the dots as greater positive voltage is applied to gate
$M$, modifying the electrostatic coupling between the dots.


We have presented a Si:P double-dot system, fabricated by ion
implantation, with gate tuneable interdot coupling. The
implementation of a double-dot system in silicon will enable
relatively straightforward incorporation into more complex
circuits. Future work on the silicon double-dots presented here,
will examine application in quantum cellular automata.
Additionally, a significant reduction in the size dots may allow
individual quantum states to be investigated. The realization of a
quantum double-dot may provide a pathway to coherent measurements
of spin states in silicon, as has been demonstrated in AlGaAs/GaAs
heterostructures \cite{petta2005}.


The authors would like to thank J.C. McCallum, M. Lay, C.I. Pakes,
S. Prawer for helpful discussions and E. Gauja, R. P. Starrett, D.
Barber, G. Tamanyan and R. Szymanski for their technical support.
This work was supported by the Australian Research Council, the
Australian Government and by the US National Security Agency
(NSA), Advanced Research and Development Activity (ARDA) and the
Army Research Office (ARO) under contract number DAAD19-01-1-0653.

\clearpage

\bibliography{dcwl_sub}

\clearpage

\begin{table}
  \centering
\begin{tabular}{|c|c|c|}
  \hline
  Transition & ~$\triangle$q$_{\mathrm{SETL}}$~/~e~ & ~$\triangle$q$_{\mathrm{SETR}}$~/~e~ \\
  \hline
  ~DOT$_{\mathrm{L}}$ $\leftrightarrow$ LEAD$_{\mathrm{L}}$~ & $\sim$ 0.074 & $\sim$ 0.043 \\
  ~DOT$_{\mathrm{R}}$ $\leftrightarrow$ LEAD$_{\mathrm{R}}$~ & $\sim$ 0.042 & $\sim$ 0.074 \\
  ~DOT$_{\mathrm{L}}$ $\leftrightarrow$ DOT$_{\mathrm{R}}$~ & $\sim$ 0.060 & $\sim$ 0.062 \\
  \hline
\end{tabular}
 \caption{Charge induced on SET$_{\mathrm{L(R)}}$ for various charge state
transitions in the double-dot system.} \label{table:one}
\end{table}

\clearpage

FIG. \ref{fig:one}: (a) Cross-section schematic of the device
showing the phosphorus implanted area, surface gates and SETs. (b)
SEM of the device, with implanted double-dot and leads, control
gates and SETs. The red shaded areas indicate regions of
phosphorus ion implantation.\\

FIG. \ref{fig:two}: (a) and (b) Simultaneously measured
conductances signals (with a best-fit plane subtracted for
clarity) from SET$_{\mathrm{L}}$ and SET$_{\mathrm{R}}$
respectively as a function of V$_{\mathrm{L}}$ and
V$_{\mathrm{R}}$ (V$_{\mathrm{M}}$ = 0 V). A double-dot charge
stability diagram is clearly resolved. Single traces are taken
from (a) and SET$_{\mathrm{L}}$ conductance is plotted as a
function of gate $L$ (c) and gate $R$ (d). The charge occupancies
of DOT$_{\mathrm{L}}$ and DOT$_{\mathrm{R}}$ are given by $m$ and
$n$, respectively. SET V$_{\mathrm{ac}}$ = 20 $\mu$V, $T$ = 50 mK
and $B$ = 0 T.\\

FIG. \ref{fig:three}: (a) SET$_{\mathrm{L}}$ transconductance as a
function of V$_{\mathrm{L}}$ and V$_{\mathrm{R}}$
(V$_{\mathrm{M}}$ = 0 V). The diagonal line follows an interdot
charge state transition. (b) SET$_{\mathrm{L}}$ conductance
(background line subtracted) as V$_{\mathrm{L}}$ and
V$_{\mathrm{R}}$ are varied along an interdot transition ($m$ + 1,
$n$) $\leftrightarrow$ ($m$, $n$ + 1). SET$_{\mathrm{L}}$
conductance as a function of V$_{\mathrm{L}}$ and V$_{\mathrm{R}}$
when (c) V$_{\mathrm{M}}$ = + 0.81 V, and (d) V$_{\mathrm{M}}$ = +
1.0 V. (e) Interdot coupling ratios C$_{\mathrm{m}}$/C$_{\Sigma
L(R)}$ as a function of V$_{\mathrm{M}}$. SET V$_{\mathrm{ac}}$ =
20 $\mu$V, $T$ = 50 mK and $B$ = 0 T.

\clearpage

\clearpage

\clearpage

\end{document}